\documentclass[10pt,twocolumn]{article}
\usepackage{authblk}
\usepackage[margin=2cm]{geometry}
\usepackage{amsmath}
\usepackage{cite}
\usepackage{url}
\usepackage{graphicx}
\usepackage{color}
\usepackage{siunitx}
\usepackage[utf8]{inputenc}
\usepackage{amssymb,subfig,bm}

\title{Learning Magnitude Distribution of Sound Fields via Conditioned Autoencoder}

\author[1]{Shoichi Koyama}
\author[2]{Kenji Ishizuka}
\affil[1]{National Institute of Informatics, Tokyo, Japan}
\affil[2]{Yamaha Corporation, Shizuoka, Japan}

\date{}

\DeclareMathOperator*{\minimize}{minimize}

\begin{document}

\maketitle
\begin{abstract}
A learning-based method for estimating the magnitude distribution of sound fields from spatially sparse measurements is proposed. Estimating the magnitude distribution of acoustic transfer function (ATF) is useful when phase measurements are unreliable or inaccessible and has a wide range of applications related to spatial audio. We propose a neural-network-based method for the ATF magnitude estimation. The key feature of our network architecture is the input and output layers conditioned on source and receiver positions and frequency and the aggregation module of latent variables, which can be interpreted as an autoencoder-based extension of the basis expansion of the sound field. Numerical simulation results indicated that the ATF magnitude is accurately estimated with a small number of receivers by our proposed method. 
%A learning-based method for estimating the magnitude distribution of sound fields is proposed. Estimating a spatial sound from microphone measurements has a wide range of applications. In particular, when microphone measurements are not taken simultaneously, or it is impossible to output a reference signal to the sound source, the target of the estimation becomes the magnitude distribution of the sound field. In such situations, it is difficult to estimate using the physical properties of the sound field because the phase distribution is unreliable or inaccessible. We propose a neural network-based method to solve this problem. The key feature of our network architecture is the input and output layers conditioned on source and microphone positions and frequency and the aggregation module of latent variables, which can be interpreted as an autoencoder-based extension of the basis expansion of the sound field. Our proposed method is evaluated by numerical simulations. 
\end{abstract}
%\keywords{\textit{acoustic transfer function, deep neural networks, autoencoder, magnitude distribution, spatial interpolation}}
%

\section{Introduction}\label{sec:introduction}

Sound field estimation, interpolation, capturing, or reconstruction is a fundamental problem in acoustic signal processing and machine learning, which aim to estimate a spatial distribution of acoustic field from a discrete set of sensor observations. The estimation of room impulse response (RIR) and acoustic transfer function (ATF, the frequency-domain representation of RIR) is a special case of the sound field estimation problem, where RIRs or ATFs between unknown sources (loudspeakers) and receivers (microphones) are estimated from those between known sources and receivers in a room. Such estimation allows us to know the acoustic characteristics without measurements and can be applied to various techniques using RIRs and ATFs, such as speech enhancement, visualization, and spatial audio. 

There have been many studies on the estimation of RIRs and complex-valued ATFs. One of the most widely used techniques is the basis-expansion-based method, which is based on the expansion of ATF into plane wave functions, spherical wave functions, or equivalent sources~\cite{Williams:FourierAcoust,Poletti:J_AES_2005,Ueno:FnT_SP2025}. The method based on kernel ridge regression with the constraint that the estimated function satisfies the Helmholtz equation is a generalization of the basis-expansion-based method to infinite-dimensional expansion~\cite{Ueno:IEEE_SPL2018,Ueno:IEEE_J_SP2021}. Recently, the learning-based method, especially the neural-network-based (NN-based) method, has attracted attention because it allows us to estimate the sound field with extremely few microphones, compared with the methods not using training data~\cite{Lluis:JASA2020,Luo:NIPS2022,Koyama:IEEE_M_SP2025}. 

In contrast to many ATF estimation methods, which target the estimation of complex amplitude in the frequency domain, this study deals with the problem of estimating the ATF magnitude, i.e., the absolute value of the complex amplitude in the frequency domain. The ATF magnitude estimation technique will be useful when the reference signal for RIR measurement from the source is unreliable or cannot be output in the first plane. Specifically, this is the case when the signals of each receiver are not synchronized or when estimating the directivity of musical instruments or other vibrating bodies. 
%It is also expected that it will be more tractable for real-valued magnitudes to be represented by NNs in the learning-based method than for complex-valued amplitudes. 
However, unlike the methods using the complex amplitudes, it is difficult for the ATF magnitude estimation to incorporate the physical properties of the sound field because the governing equation of magnitude distribution cannot be well represented. 

We propose a learning-based method based on an autoencoder conditioned on the source position, receiver position, and frequency for estimating the ATF magnitude distribution. In many current learning-based methods, ATFs at fixed target positions are estimated from ATFs at observation positions as their subset~\cite{Lluis:JASA2020,Miotello:ICASSP2024}. Therefore, it is difficult to estimate ATF at arbitrary positions and to combine multiple datasets with different measurement setups. Although there are no such restrictions with the method based on implicit neural representation or neural field (NF)~\cite{Sitzmann:NeurIPS2020,Mildenhall:ACM_Commun2021}, the network needs to be retrained during inference, which is computationally intensive. Our approach, based on the conditioned autoencoder, overcomes these shortcomings. Our proposed method can be considered a nonlinear extension of the basis-expansion-based method and is an extension of the method we proposed for estimating the magnitude of head-related transfer function (HRTF)~\cite{Ito:IWAENC2022}. 

\section{Problem Statement}

ATF from the source position $\bm{y}\in\mathbb{R}^3$ to the receiver position $\bm{x}\in\mathbb{R}^3$ at the angular frequency $\omega\in\mathbb{R}$ is denoted as $h(\bm{x}, \bm{y}, \omega)$ (i.e., $h: \mathbb{R}^3 \times \mathbb{R}^3 \times \mathbb{R} \to \mathbb{C}$). Suppose that ATF magnitude in logarithmic scale is denoted as $a(\bm{x}, \bm{y}, \omega):=20 \log_{10} |h(\bm{x}, \bm{y}, \omega)|$, and it is measured at multiple positions, which are sparsely distributed inside the target region $\Omega \subset \mathbb{R}^3$. Our goal is to estimate $a$ of the given source position $\bm{y}$ at the target positions $\{\bm{x}_n^{(\mathrm{t})}\}_{n=1}^{N}$ inside $\Omega$ from its sparse measurements at the measurement positions $\{\bm{x}_m^{(\mathrm{m})}\}_{m=1}^M$ ($M \ll N$). 

In practice, the log-ATF magnitude is obtained for multiple source positions and discrete frequency bins. By defining the indexes of sources and frequencies as $l\in\{1,\ldots,L\}$ and $f\in\{1,\ldots,F\}$, respectively, the discrete value of the log-ATF magnitude is denoted as $a_{m,l,f}$ for the measurement positions and $a_{n,l,f}$ for the target positions, where $L$ and $F$ are the number of sources and frequency bins, respectively. 

\section{Prior Work}

We here briefly introduce prior work on RIR and ATF estimation. Many previous studies focus on the estimation of the spatial distribution of RIR or complex-valued ATF. There are few studies on the estimation of magnitude distribution, with a few exceptions~\cite{Lluis:JASA2020,Liang:EURASIP2024,Miotello:ICASSP2024}. 

%The spatial interpolation problem of RIR and complex-valued ATF has been a fundamental problem in acoustic signal processing because it has a wide range of applications related to spatial audio~\cite{Poletti:J_AES_2005,Ueno:FnT_SP2025}. One of the most widely used techniques is the basis-expansion-based method. In recent years, machine-learning-based methods, especially neural-network-based methods, have been intensively investigated~\cite{Koyama:IEEE_M_SP2025}. In contrast, there are few studies on spatial interpolation for magnitude distribution, with a few exceptions~\cite{Lluis:JASA2020,Liang:EURASIP2024,Miotello:ICASSP2024}. 

\subsection{Basis-expansion-based methods} \label{sec:basis-exp}

In the basis-expansion-based methods, the ATF complex amplitude $h(\bm{x},\bm{y},\omega)$ is approximated as the linear combination of basis function $\varphi_j$ as 
\begin{align}
h(\bm{x},\bm{y},\omega) \approx \sum_{j=1}^J \gamma_j (\bm{y},\omega) \varphi_j (\bm{x},\bm{y},\omega),
\label{eq:basis-exp}
\end{align}
where $j$ is the index, $J$ is the number of basis functions, and $\gamma_j$ is the $j$th expansion coefficient. For the basis function $\varphi_j$, plane wave function, spherical wave function, and equivalent source are typically used because these functions are the element solution of the Helmholtz equation~\cite{Williams:FourierAcoust,Poletti:J_AES_2005,Ueno:FnT_SP2025}. For the measurement positions $\{\bm{x}_m^{(\mathrm{m})}\}_{m=1}^M$, Eq.~\eqref{eq:basis-exp} is rewritten as 
\begin{align}
\bm{h} \approx \bm{\Phi} \bm{\gamma},
\end{align}
where $\bm{h}\in\mathbb{C}^M$, $\bm{\Phi}\in\mathbb{C}^{M \times J}$, and $\bm{\gamma}\in\mathbb{C}^J$ consist of $\{h(\bm{x}_m^{(\mathrm{m})}, \bm{y}, \omega)\}_{m=1}^M$, $\{\varphi_j(\bm{y},\omega)\}_{j=1}^J$, and $\{\gamma_j(\bm{y},\omega)\}_{j=1}^J$, respectively. 

The expansion coefficients $\{\gamma_j\}_{j=1}^J$ are generally estimated by solving the following linear regression problem:
\begin{align}
\minimize_{\bm{\gamma}\in\mathbb{C}^J} \left\| \bm{h} - \bm{\Phi}\bm{\gamma} \right\|^2 + \lambda \| \bm{\gamma} \|_2^2,
\end{align}
where $\lambda \in \mathbb{R}_{\ge 0}$ is the regularization parameter. The expansion coefficients are simply obtained as
\begin{align}
\hat{\bm{\gamma}} = \left( \bm{\Phi}^{\mathsf{H}} \bm{\Phi} + \lambda \bm{I} \right)^{-1} \bm{\Phi}^{\mathsf{H}} \bm{h}.
\end{align}
The ATF at the target positions can be obtained by substituting the estimated expansion coefficients into Eq.~\eqref{eq:basis-exp}.

Kernel ridge regression is also frequently used in the sound field estimation problem, and it can be seen as a generalization of the basis-expansion-based methods to infinite-dimensional expansion. In \cite{Ueno:IEEE_SPL2018,Ueno:IEEE_J_SP2021}, the kernel ridge regression with the constraint that the estimated function satisfies the Helmholtz equation has been proposed.

\subsection{Neural-network-based methods}

Deep neural networks (DNNs) have also been applied to the sound field estimation problem because of their high adaptability and representational power~\cite{Lluis:JASA2020,Luo:NIPS2022,Koyama:IEEE_M_SP2025}. There are several approaches to apply DNNs to the ATF estimation problem. One approach is to set the DNN output to discretized values of ATF at the target positions~\cite{Lluis:JASA2020,Pezzoli:Sensors2022}. The DNN input is the observations at the measurement positions chosen from the target positions. Thus, the DNN parameters are optimized by minimizing a loss function defined for a pair of observation and target ATF data. 

The other approach is the use of implicit neural representation or NF, where the continuous function $h$ is represented by DNNs~\cite{Pezzoli:ForumAcusticum2023,Ribeiro:IEEE_ACM_J_ASLP2024}. The input and output of DNN are the argument of $h$, i.e., $\bm{x}$, and the scalar function value approximating $h(\bm{x})\in\mathbb{C}$. The implicit neural representation is typically used in PINNs by incorporating the regularization term that evaluates the deviation from the governing equation~\cite{Karniadakis:NatRevPhys2021,Raissi:CompPhys2019}. 

The main drawback of the first approach is the predefined target positions. Since the DNN model is trained to estimate the ATFs at the fixed target positions from the measurements at their subset positions, it is difficult to combine multiple datasets measured on different setups. The second approach, the NF-based approach, makes it possible to estimate the ATF at an arbitrary target position. However, the NF-based methods are usually applied to estimate the ATF only with a single observation. To apply the NF-based method as a learning-based method, the NF-based DNN model is trained by using the training data, and the DNN parameters are optimized again for the test data, which generally requires high computational cost.  

\section{Proposed Method}

\begin{figure*}[ht]
 \centerline{
 \includegraphics[width=2.0\columnwidth]{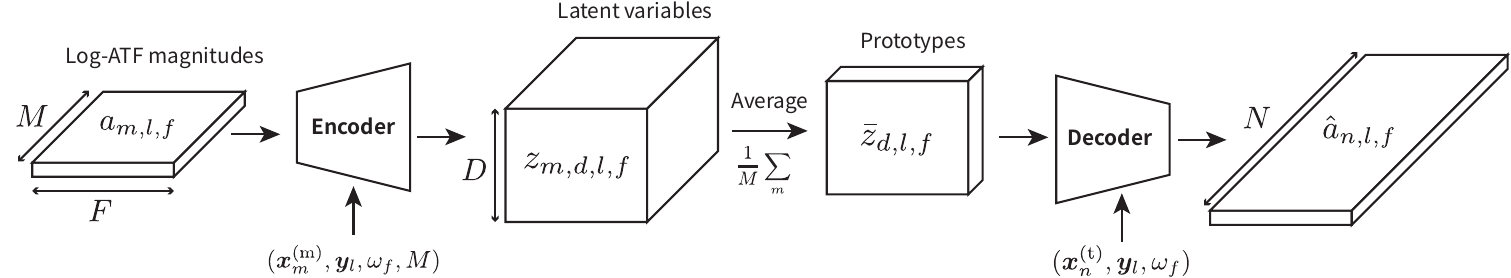}}
 \caption{Outline of proposed network architecture.}
 \label{fig:nn}
\end{figure*}

In the basis-expansion-based methods described in Section~\ref{sec:basis-exp}, the sound field is expanded by predefined spatial basis functions, and the expansion coefficients are obtained by linear regression using the measurements. The expansion coefficients are generally independent of the receiver position, whereas the basis functions depend on the receiver position. 

We propose a learning-based method for estimating the log-ATF magnitude, which is based on the DNN trained by $a_{n,l,f}$ for multiple source positions. Our proposed method is regarded as a nonlinear extension of the basis-expansion-based method by using an autoencoder consisting of the encoder and decoder conditioned on the receiver positions and the receiver-position-independent latent variables, i.e., the input of the decoder. A similar technique has been applied to the HRTF upsampling~\cite{Ito:IWAENC2022}. 

\subsection{Model}

Figure~\ref{fig:nn} shows the proposed network architecture, which consists of an encoder, an aggregation module, and a decoder. Based on the insights described above, the encoder and decoder weights depend on receiver positions, whereas the decoder input is independent of receiver positions. In addition, the encoder is conditioned on the source positions, frequency, and the number of observations, and the decoder is conditioned on the source positions and frequency. 

The detailed network architecture is almost the same as the one proposed in~\cite{Ito:IWAENC2022}. The weights and biases of the encoder and decoder are independently generated by NN with the input of the conditioning vector. The conditioning vector is a concatenation of the receiver positions, source positions, angular frequencies, and number of observations (only for the encoder). The elements of the conditioning vector are normalized by their maximum values. Then, Fourier feature mapping (FFM)~\cite{Tancik:NeurIPS2020} is applied as a preprocessing step. The weight/bias generators consist of fully connected layers, layer normalization, and Mish nonlinearity. 

The encoder transforms the sparse log-ATF magnitudes $a_{m,l,f}$ into latent variables $z_{m,d,l,f}$, which consists of layer normalization, the Mish nonlinearity, and hyper-linear layers~\cite{Ha:ICLR2017}. Then, the latent variables, $z_{m,d,l,f}$, are averaged over the measurement positions as
\begin{align}
\bar{z}_{d,l,f} = \frac{1}{M} \sum_{m=1}^M z_{m,d,l,f},
\end{align}
which is referred to as prototypes as used in the prototypical network~\cite{Snell:NeurIPS2017}.

The decoder generates the log-ATF magnitudes $\hat{a}_{n,l,f}$ at the target positions from the prototypes $\bar{z}_{d,l,f}$, which consists of layer normalization, the Mish nonlinearity, and hyper-linear layers.

\subsection{Loss function}

The loss function $\mathcal{L}$ for the training is defined as the mean log-spectral distortion (LSD) of the ATF as
\begin{align}
\mathcal{L} = \frac{1}{NL} \sum_{n,l} \sqrt{\frac{1}{F} \sum_f \left( \hat{a}_{n,l,f} - a_{n,l,f} \right)^2}.
\end{align}

\section{Experimental evaluations}

\begin{figure}[t]
 \centerline{
 \includegraphics[width=0.7\columnwidth]{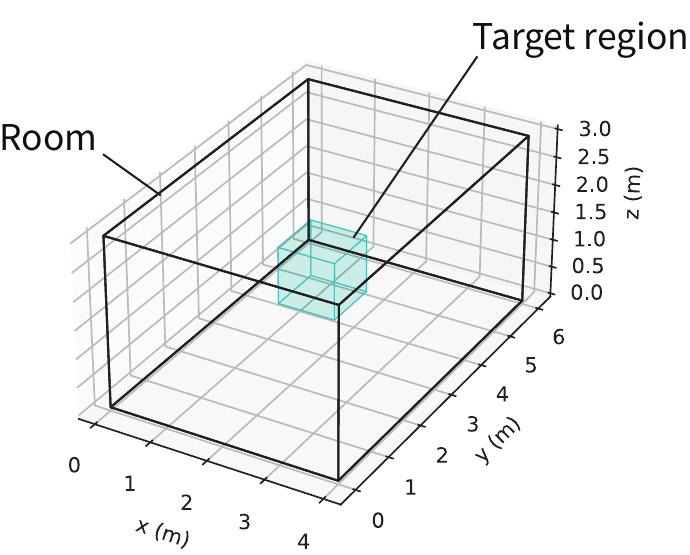}}
 \caption{Experimental setting.}
 \label{fig:geometry}
\end{figure}

We evaluated the proposed method, comparing it with the method based on the kernel ridge regression and NF-based method. 

\subsection{Experimental setup}

\begin{figure}[t]
 \centerline{
 \includegraphics[width=1.0\columnwidth]{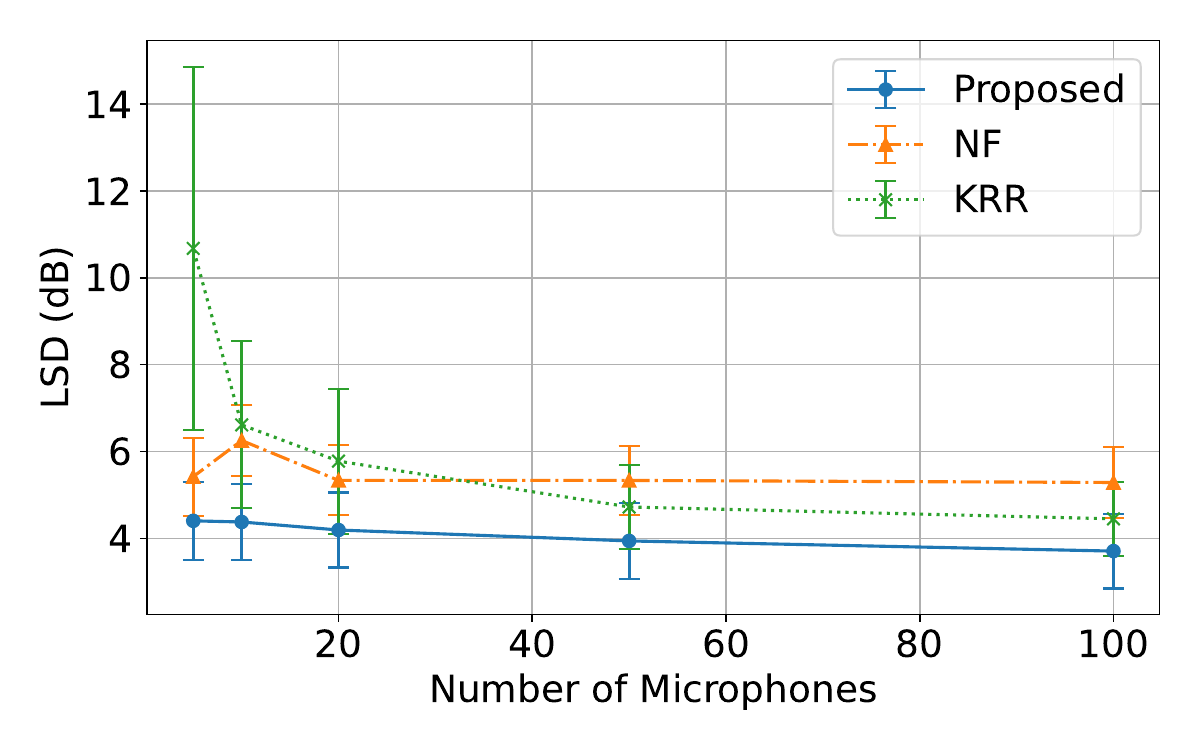}}
 \caption{Average LSD with respect to the number of microphones.}
 \label{fig:LSD}
\end{figure}

We generated an ATF dataset of a single room by using acoustic simulation based on the image source method~\cite{Allen:JASA1979}. As shown in Fig.~\ref{fig:geometry}, the shoebox-shaped room with the size of $4.0~\mathrm{m} \times 6.0~\mathrm{m} \times 3.0~\mathrm{m}$ was assumed. The cuboid target region was set at the center of the room, and its size was $1.0~\mathrm{m} \times 1.0~\mathrm{m} \times 1.0~\mathrm{m}$. The 1331 target positions were obtained by discretizing the target region every $0.1~\mathrm{m}$. The 1024 source positions were randomly chosen within the room, excluding the target region. The RIRs were generated by setting the target reverberation time to $200~\mathrm{ms}$ by using pyroomacoustics~\cite{pyroomacoustics}. The sampling frequency was $2000~\mathrm{Hz}$, so the maximum frequency of the target was $1000~\mathrm{Hz}$. The log-ATF magnitudes were obtained by truncating the RIR into $128$~samples and Fourier transform. The ATF dataset of the 1024 source positions was separated into training, validation, and test data, which contain 820, 122, and 122 log-ATF magnitudes, respectively. We investigated $M=5$, $10$, $20$, and $100$ for the number of measurements, and the measurement positions were randomly chosen from the target positions. 

The proposed DNN was trained for 1400 epochs using Adam optimizer~\cite{Kingma:ICLR2015}. The learning rate was $10^{-3}$ for the first 800 epochs, $10^{-4}$ for 801-1200 epochs, and $10^{-5}$ for 1201-1400 epochs. The number of measurements $M$ used for the training was larger than that for the validation and test. For example, the network for $M=5$ was trained for $M=5$, $10$, $20$, and $100$, but that for $M=100$ was trained only for $100$. The used DNN for evaluation was at the epoch with the smallest validation loss. 

In the kernel ridge regression (KRR), the log-ATF magnitudes at the target positions were estimated from the observations only in the test data because KRR is not a learning-based method. For the kernel function, we used Gaussian kernel~\cite{Murphy:ML} because the constraint of the Helmholtz equation is not valid for the magnitude measurements. The precision (the inverse of the variance) in the Gaussian kernel was set to $10^{-2}$. The regularization parameter in KRR was set to $10^{-3}$.

The DNN model for the NF-based method had the input of the source position, receiver position, and angular frequency and the output of the log-ATF magnitude. The input was normalized by their maximum values, and FFM was applied. Then, three fully connected layers of 128 neurons and the ReLu activation function~\cite{Agarap:arxiv2018} were applied. The loss function was the LSD of the ATF magnitude. The network was trained on the training data with 1400 epochs and validated and tested by fine-tuning the entire network with 10 epochs, using Adam optimizer. The learning rate was $10^{-5}$.

\subsection{Experimental results}

\begin{figure}[t]
 \subfloat[Proposed]{\includegraphics[width=1.0\columnwidth]{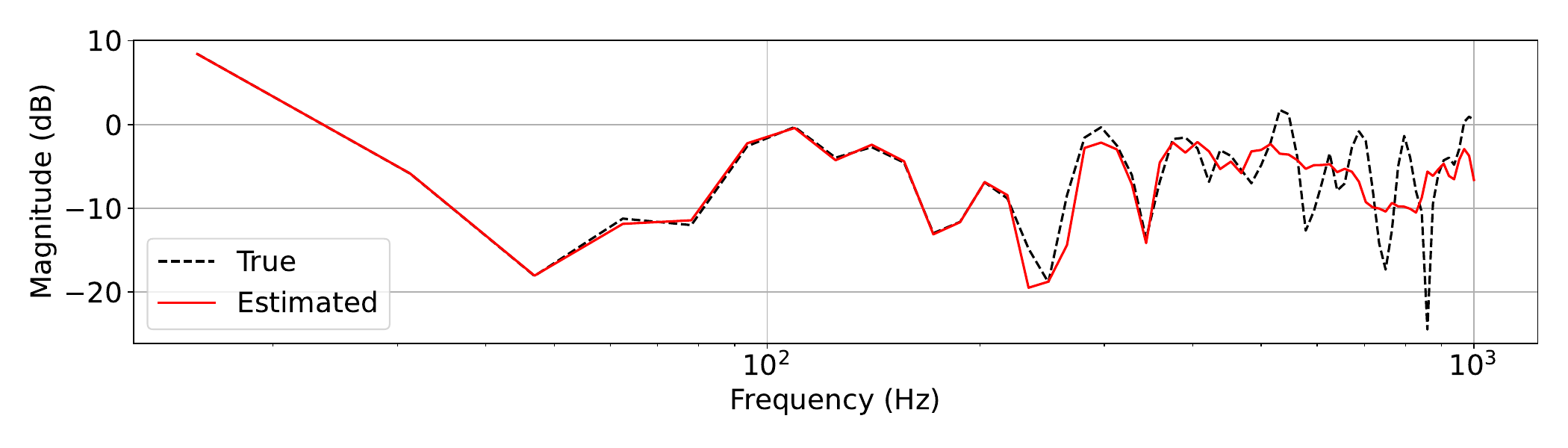}}\\
 \subfloat[NF]{\includegraphics[width=1.0\columnwidth]{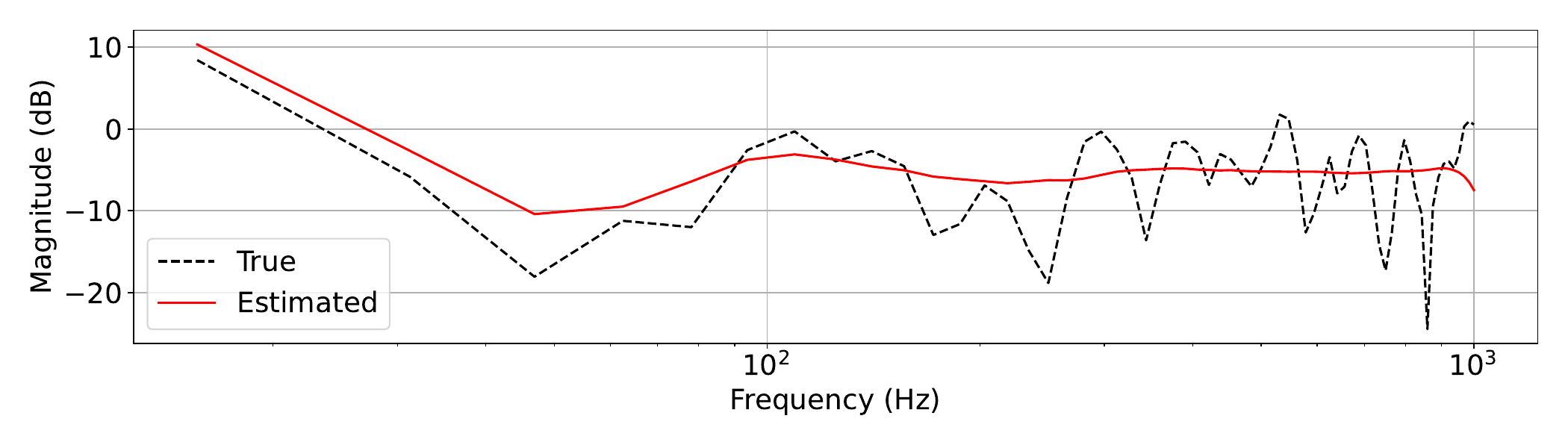}}\\
 \subfloat[KRR]{\includegraphics[width=1.0\columnwidth]{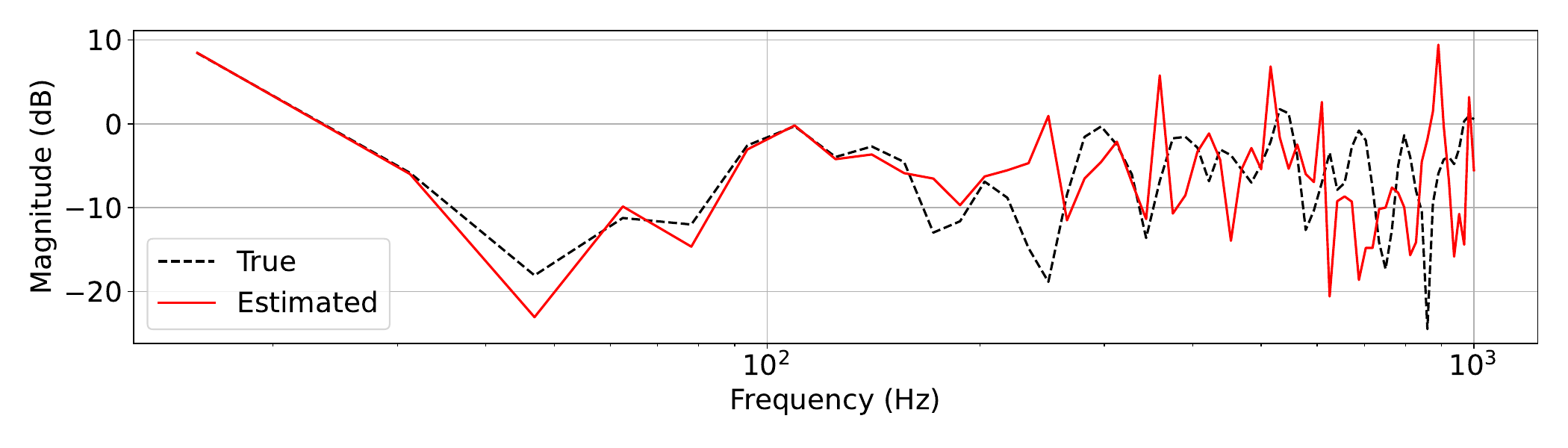}}
 \caption{Estimated ATF magnitude at the center of the target region when $M=5$.}
 \label{fig:atf_mag}
\end{figure}

\begin{figure}[t]
 \subfloat[Ground truth]{\includegraphics[width=0.5\columnwidth]{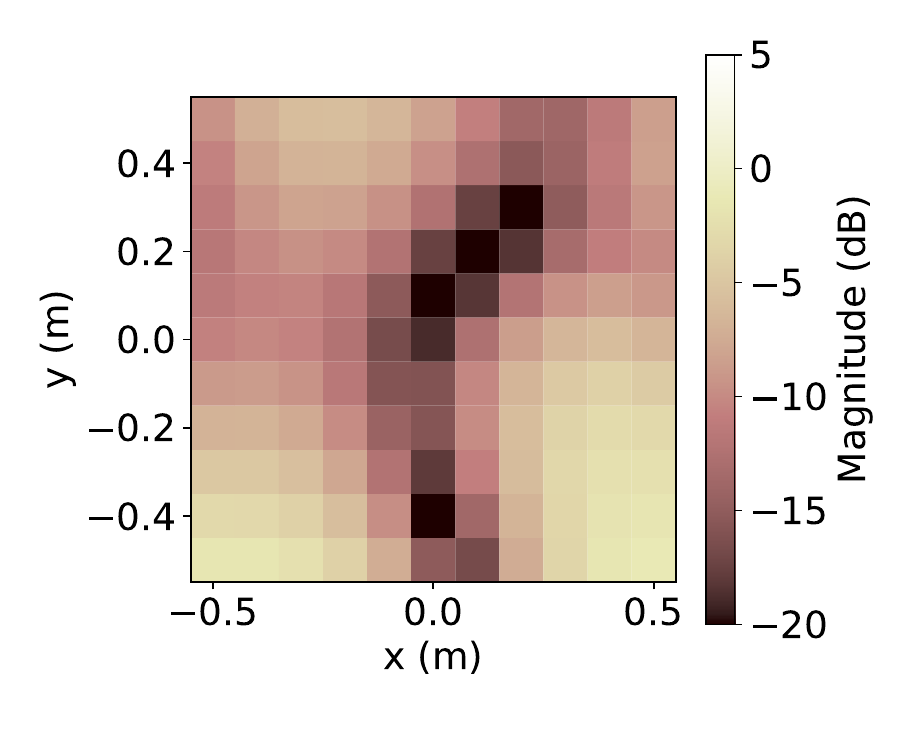}}
 \subfloat[Proposed]{\includegraphics[width=0.5\columnwidth]{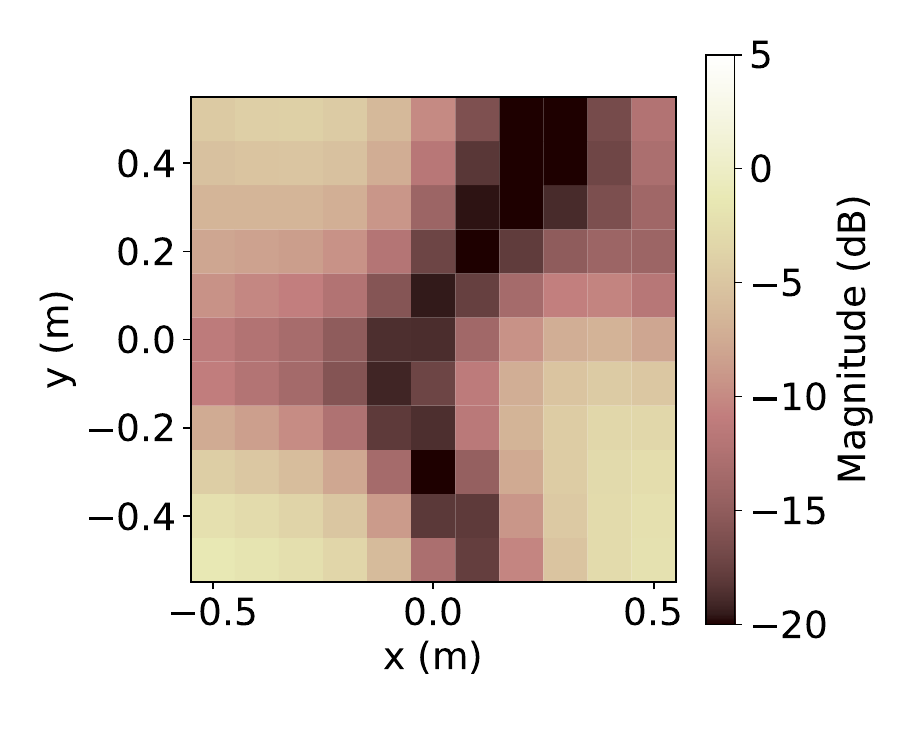}}\\
 \subfloat[NF]{\includegraphics[width=0.5\columnwidth]{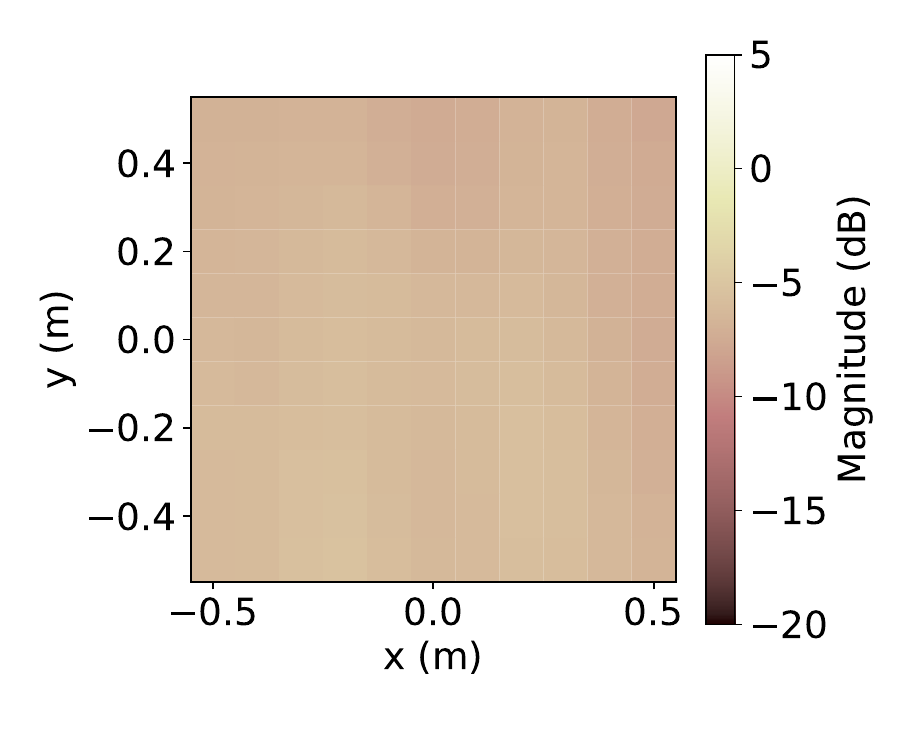}}
 \subfloat[KRR]{\includegraphics[width=0.5\columnwidth]{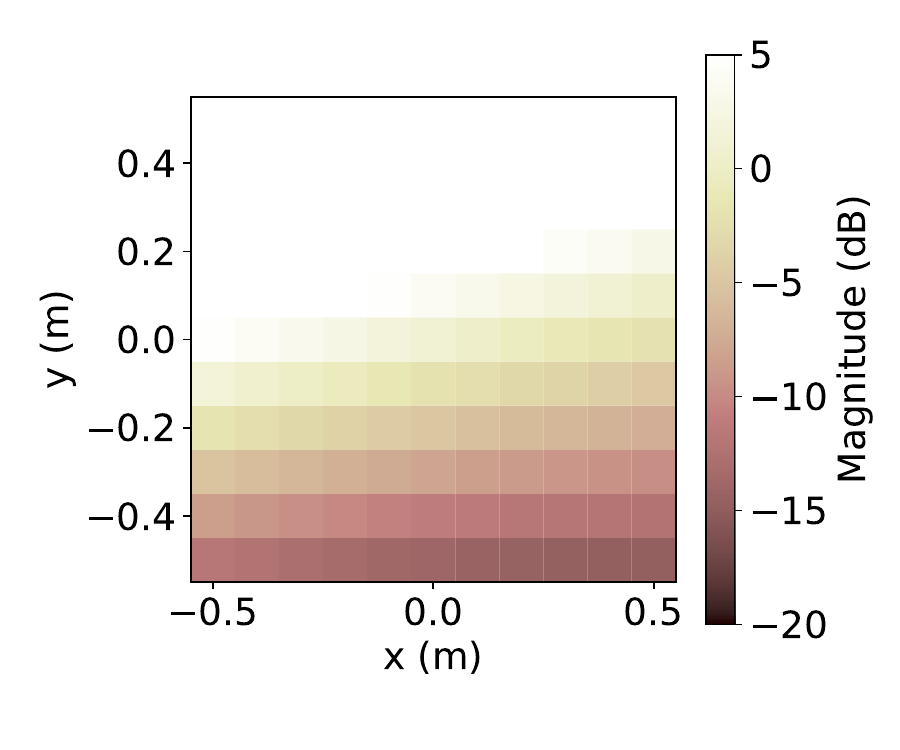}}
 \caption{Magnitude distribution on $x$-$y$-plane at $z=0$ at $250~\mathrm{Hz}$ when $M=5$.}
 \label{fig:dist}
\end{figure}

Figure~\ref{fig:LSD} shows the average LSD with the standard deviation on the test data with respect to the number of measurements $M$. The LSD of KRR deteriorates rapidly as the number of observations decreases because KRR does not use the training data. The NF-based method showed nearly constant LSDs. The LSDs of NF were lower than those of KRR for $M=5$, $10$, and $20$, but higher for $M=50$ and $100$. The LSDs of the proposed method were the lowest for all the numbers of observations. Therefore, the learning-based methods are effective when the number of observations is small, owing to the extraction of ATF properties from the training data. Since the proposed method showed high performance in both cases where the number of observations is small and large, the proposed model can successfully learn the characteristics of ATF.

The true and estimated ATF magnitudes at the center of the target region for the case of $M=5$ are shown in Fig.~\ref{fig:atf_mag}. The estimated ATF of KRR fluctuated to a large extent, especially above $300~\mathrm{Hz}$. The ATF magnitude of NF was like a smoothed version of the true ATF. Although the LSD was smaller than that of KRR, it did not capture the fine structure of ATF at all. In contrast, the proposed method captured fine structures up to approximately $400~\mathrm{Hz}$ and also captured the general shape at frequencies above that. 

Figure~\ref{fig:dist} shows the spatial distributions of the true and estimated ATF magnitude on the $x$-$y$-plane at $z=0$ in the case of the frequency $250~\mathrm{Hz}$ and $M=5$. The true distribution shows large peaks and dips, but the estimated distribution of KRR was slanted largely in the positive $y$ direction, while that of NF was almost constant. In the proposed method, although the estimated distribution was somewhat different from the true distribution in the region of $x>0~\mathrm{m}$ and $y>0.2~\mathrm{m}$, it showed a magnitude distribution similar to the true one. 

\section{Conclusion}

We proposed a method for estimating the ATF magnitude distribution based on autoencoder conditioned on source and receiver positions and frequency. Our proposed method allows us to combine multiple datasets measured by different measurement setups and also keep the computational complexity low during inference. Numerical experiments have shown that the proposed method achieves high estimation accuracy even when there are only a few observations. 

\section{Ackowledgments}
This work was supported by JSPS KAKENHI Grant Number 23K24864. 

% For bibtex users:
\bibliographystyle{ieeetr}
%\bibliography{st_def_abrv,skoyamalab_en,refs}
\bibliography{skoyamalab_en,refs}

\end{document}